**Title**: How Honeybees Perceive and Traverse Apertures

**Authors**: T. Jakobi[1], M. Garratt[1], M. Srinivasan[2] and S. Ravi[1]

**Affiliations:**

[1] School of Engineering and Technology, University of New South Wales Canberra,
ACT 2600, Australia

[2] Queensland Brain Institute, University of Queensland, St. Lucia, QLD 4072, Australia

**Corresponding Author's Email:** timrjak@gmail.com, sridhar.r@unsw.edu.au

**Abstract**

The ability to fly through openings in vegetation allows insects like bees to access otherwise unreachable food sources. The specific visual strategies employed by flying insects during aperture negotiation tasks remain unknown. In this study, we investigated the visual and geometric parameters of apertures that influence traversing honeybees. We recorded honeybees flying through apertures with varying shapes and sizes using high-speed cameras to examine their spatial distribution patterns and trajectories during passage. Our results reveal that the flight of bees was, on average, along the bilateral center of the edges of the aperture irrespective of the size. When apertures were smaller, bees tended to also fly closer to the vertical center. However, for larger apertures, they traversed at lower vertical positions (closer to the bottom edge). The behaviors suggest that honeybees modulate their flight trajectories in response to spatial constraints, adjusting trajectory relative to aperture dimensions. When entering at off-center horizontal positions, bees tended to access the vertical center of the aperture, indicating altitude selection influenced by the curvature of the edge below. This behavior suggests an acute awareness of the vertical and horizontal spatial constraints and a preference for maintaining a curvature-dependent altitude that optimizes safe passage. Our analysis reveals that honeybees modulate speed and altitude above the ventral edge passing beneath them, maintaining a ventral optic flow magnitude within a preferred range. This relationship suggests a control mechanism where bees rely on visual information in a narrow ventrally directed field to navigate safely through confined spaces.

**Summary Statement:** This study identifies ventral optic flow cues that honeybees use to navigate apertures of varying shapes and sizes, using visual cues to adjust flight paths for safe and efficient passage.

## Introduction

Flying through dense natural environments demands agility and precise coordination, as flyers in these situations must continuously adapt to complex visual and spatial cues to avoid obstacles and find safe passage (Srinivasan, 2011; Serres and Ruffier, 2017). In both natural and urban landscapes, apertures—gaps between foliage or artificial structures—frequently pose navigational challenges, driving the evolution of sensory and motor responses that enable organisms to perform routine foraging, evade predators, and select suitable habitats (Norberg, 2012).

Despite their small size and limited neural resources, flying insects have developed efficient visual processing systems that enable sophisticated navigation in complex environments (Egelhaaf and Kern, 2002). Their compound eyes, which are effectively monocular, provide a wide field of view but offer limited spatial resolution and depth perception (Horridge, 1977). To navigate effectively, bees organize and process optic flow (OF)—the pattern of motion in the visual environment induced by their own movement—from different regions of their eyes: lateral, ventral, dorsal, and frontal, each playing a distinct role in flight control (Gibson, 1950; Srinivasan et al., 1996; Portelli et al., 2011). The ventral region is particularly crucial for maintaining flight speed and altitude, with ventral optic flow playing a key role in speed control, landing, and stabilization (Srinivasan et al., 1991; Izzo and De Croon, 2012; Serres and Ruffier, 2017). Lateral regions assist in horizontal positioning within passages, and bees use lateral landmarks for frontal target acquisition (Lehrer, 1990). The frontal and dorsal regions, with higher visual acuity due to smaller interommatidial angles, facilitate obstacle detection and avoidance (Land, 1997). By segmenting visual input across these specific retinal areas, bees effectively respond to information pertinent to their immediate navigational needs, synthesizing complex spatial information into actionable cues that help regulate flight and navigation through complex environments.

During forward flight, ranging of peripheral objects in the lateral visual field is facilitated by the magnitude of lateral optic flow; closer side objects produce stronger translational optic flow signals, enabling the flying agent to adjust its position relative to these objects (Gibson, 1950; Srinivasan et al., 1996). To facilitate collision-free flight through narrow passages, insects balance the magnitudes of lateral optic flow on both sides, adjusting their trajectory to maintain symmetry and avoid collisions (Tammero and Dickinson, 2002). Thus, amid confined regions of clutter, flying insects reduce their speed, which constrains the availability of inferable spatial cues (Serres and Ruffier 2017). Insects have also been shown to use expansional optic flow—characterised by the focus of expansion for behaviours such as landing and for detecting imminent collisions with vertical surfaces (Srinivasan et al., 1991; Schrater et al., 2001). Recent studies have reported evidence that insects estimate the time-to-contact ($\tau$) or a closely related metric for collision avoidance (Balebail et al., 2019; Ravi et al., 2022). In the context of passing through apertures, expansional optic flow may not provide the precise cues required for centering and determining distance from edges. In contrast, translational optic flow offers information for adjusting lateral position, vertical position, and speed during such manoeuvres.

The detection of large apertures and the control of flight through them can also be mediated by brightness-based cues if the aperture is brighter (or darker) than the surrounding surface (Baird and Dacke, 2016). In the case of smaller apertures, flying insects enhance navigation and spatial perception through their antennae. Antennae provide airflow mechanosensory signals that enable faster speed estimations compared to relying solely on vision (Fuller et al., 2014). Terrestrial bees have been observed mapping the spatial properties of objects with their antennae (Erber et al., 1997), suggesting that antennal probing aids in perceiving aperture size via tactile feedback. For small gaps in a wall obstruction, side-to-side probing of gap edges through reciprocal lateral gazing movements has been documented in bumblebees (Ravi et al., 2020). This involves comparing the foreground shape to the more distant background, utilizing motion parallax for spatial dimensionalization (Wallace, 1959; Sobel, 1990; Srinivasan et al., 1990; Collett, 2002). When the motion parallax cues are reduced by decreasing the displacement between the foreground and the background, bumblebees execute more vigorous scanning movements (Ravi et al., 2019), facilitating millimeter-accurate postural adjustments for gap traversal (Ravi et al., 2020). Honeybees have been shown to prefer wider apertures, suggesting that they are capable of discriminating differences in aperture width of 30% (Ong et al., 2017).

In this study, we investigated the speed and position of honeybees as they traversed apertures of varying geometries. Our goal was to determine visual cues possibly used by the bees to traverse apertures safely. By understanding this process, we aim to reveal specialized systems that flying insects use to assess and respond to spatial constraints with greater accuracy.

## Materials and methods

### Experiment setup and procedure

This study was conducted using a colony of honeybees (*Apis mellifera*) housed outdoors on a rooftop at the UNSW Canberra campus. Bees were trained to navigate to a nearby rooftop terrace (< 20 m away from the hive) equipped with an overhead awning, which filtered incident sunlight to produce uniform, diffuse illumination across the experimental area. Here, bees were further trained to locate and fly one-way through a tunnel 0.7 m high, 0.4 m wide, and 1.16 m long (Fig. 1A and Supplementary Movie 1). The tunnel was constructed from a 5 mm thick twin-wall extruded corrugated plastic sheet material (also known by its trade name Corflute). In the vertical side face at one end, the experimental setup featured a 5 cm entrance aperture (inlet), while at the opposite end a 6.5 cm diameter outlet was positioned in the ceiling. Directly beneath the ceiling outlet, a gravity feeder dispensing a homemade 10% (v/v) sucrose solution was positioned level with the height of the entrance aperture (Fig. 1A and Fig. 1D). To facilitate orientation and visual processing, the walls and floor of the tunnel were lined with a grayscale 1/f noise cloud pattern offering a mean contrast of 50%, which is consistent with visual stimuli used in related studies (Ravi et al., 2019; Monteagudo et al., 2022) (see Supplementary Movie 1). The ceiling of the tunnel was made from a transparent 80% UV-blocking acrylic sheet, allowing natural lighting from above.

Foraging bees were trained to fly through the tunnel by placing a cotton bud soaked in a 50% sucrose solution at various incremental positions along the tunnel. This approach enabled the bees to learn the location of the food source and resulted in a consistent cycle of keen bees travelling one way through the tunnel. After this initial training, the bees consistently returned to the tunnel without further conditioning. After several days of habituation, we introduced a transverse partition midway along the tunnel (57 cm away from the inlet), which served as a mounting platform for interchangeable apertures (Fig. 1A-B). After inserting the transverse wall, the setup required bees to negotiate the selected aperture in the wall to access the chamber containing the food source on the far side. To avoid potential behavioural influences from contrast variations, the transverse wall surrounding the apertures was textured with a simple black-and-white checkerboard pattern rather than the 1/f pattern used on the walls and floor (Fig. 1B). This texture was designed to provide consistent visual stimuli, reducing the likelihood of unintended visual cues affecting the bees' navigation behaviour while still providing visual texture for motion awareness.

The gravity feeder was mounted at the far end of the post-partition feeding chamber on a symmetrical platform elevated 28 cm above the floor, with the base of the feeder (bright yellow) aligned with the midline (y = 0) of the tunnel (Fig. 1A). The center of each modular aperture insert was aligned with this height along the longitudinal axis of the tunnel, ensuring that the centroid of each aperture corresponded to the Y = 0, Z = 0 (X-axis) intersection of the wall plane within the tunnel. This alignment ensured a straight line of sight from the tunnel inlet, through the aperture, and to the feeder at the opposite end, regardless of the aperture's size or shape (Fig. 1B). The camera was just below this axis of alignment, making the flight path of bees visible from aperture approach to landing on the feeder. A continuous curved wall section was installed at the far end of the tunnel (Fig. 1A) to obscure the corners from the bees' view and reduce potential visual distractions during aperture traversal.

Apertures were either circular or stadium-shaped (comprising a composite of two semi-circular arcs with a rectangular segment in between), with the stadium-shaped apertures having a minor-axis length of 60 mm and oriented either vertically or horizontally. The circular apertures had diameters ranging from 30 to 150 mm. All apertures featured the aforementioned uniform surrounding checkerboard pattern on the mounting wall; however, an additional subset of four 60 mm circular apertures was tested in which the uniform checkerboard texture was replaced by flanked textures applied exclusively to four regions – north (surrounding the semicircular half above the centre), right (surrounding the semicircular half right of the centre), left (surrounding the semicircular half left of the centre) and south (surrounding the semicircular half below the centre) (see Fig. 2).

Aperture interchanges were performed on separate days, and following each modification, bees were allowed a three-hour acclimation period before commencing data collection. Bees were not individually tagged, which is a limitation of this study as it prevents the identification of individual bees. However, the high frequency of forager activity (20 flights per minute) and the irregular nature of each test session (random days over a two-month period during summer) was considered sufficient to eliminate data biasing through repeated trials by individuals, ensuring that the recorded data reflected initial navigational responses to the aperture challenge. After a manual assessment of the recordings, only flights through the apertures that were direct, without hesitation, and free from interference by other bees in the tunnel were included as data samples. We observed a marked difference between the flights of untrained (or "new") bees and those of trained bees. Trained bees consistently made direct flights to the food source on the opposite side of the wall. It was evident that bees learned to access the food source more rapidly with larger apertures, likely due to the wider and more accessible openings. To maintain consistency in tests, the same training duration was used for all aperture changeovers. Each experimental session lasted one hour during peak foraging times, providing at least 500 recorded bee movements through each of the 11 aperture types tested.

**Data acquisition**

We positioned a GoPro Hero 11 camera just beneath the tunnel inlet, approximately 30 cm above the ground (Fig. 1A). In the GoPro settings, 'Linear Mode' was enabled to apply a preconfigured digital correction for the fisheye lens distortion characteristic of the GoPro camera. The camera view was just below the longitudinal X axis running through the centre of the apertures and feeder, facing the obstacle wall, to provide an orthogonal perspective of that plane (Fig. 1B). This setup ensured that the location of the bees' entries on the wall plane (YZ) could be accurately captured. Additionally, a mirror strip measuring 15 cm x 4 cm x 0.4 cm (L x W x T) was mounted above the aperture, with its lower edge positioned 20 cm above the X-axis to provide a time-synchronised overhead view (XY) of the bees as they navigated through the apertures within the view frame of the camera. This mirror was adjusted to a 43-degree slant (closer edge tilted up from horizontal), enabling the camera positioned at the inlet just below the X-axis to capture a top-down view of the aperture (Fig. 1B-D). A 3D coordinate system with its origin in the centre of the apertures was constructed using the frontal view for YZ coordinates and the top-down view for XY coordinates, as shown in Fig. 1A-B. With this setup, the camera captured the point where the bees intersected the plane of the wall (and aperture), and the entry time, which was used in the front-facing view for determining the entry position of the bees.

Videos were captured at a frame rate of 120 fps. Post-processing was conducted using custom MATLAB code, which involved manual image-based coordinate digitization. Spatial calibration (converting pixels to metric units) was achieved by using the known aperture dimensions in the two different views (XY and YZ) of the aperture (frontal and mirror views). To verify the X-axis calibration and determine the intersection with the transverse wall plane, using the mirror (visible in the front-facing camera view, see Fig. 1B), images of a handheld measuring scale (a ruler) held parallel to the X-axis at the centre of the apertures were captured and analysed. This allowed for identification of the intersection between the X-axis and the YZ plane, and provided a true distance measurement to convert pixels within the mirror view from the front-facing camera into metric units. The upper-, forward slanting edge of the mirror was determined to project the X = 10 cm axial position at Z = 0. Z-coordinates measured in the camera view at X = 10 cm were rescaled to the wall plane (X = 0 cm) by multiplying by the ratio of camera distance along X ($d_{wall}$) to the two measurement planes: $\frac{d_{wall}-10}{d_{wall}} = \frac{47}{57} = 0.825$. Because the vertical range of sampled bee heights is small relative to the axial camera distance, depth magnification in the mirror view was considered negligible.

Entry locations and times for each bee were identified by manually scrolling through video frames, using both the top-down mirror view (XY plane) and the axially directed frontal view (YZ plane). In addition to digitally marking the YZ entry locations (see Supplementary Movie 3), the presence of any form of collision (from the wing, body, or legs on the perimeter of the aperture) was noted, marked and categorised (see Supplementary Movie 4). In 30 entry samples for each of the apertures, the position of the distal tip of the abdomen was digitized in two consecutive frames to estimate the instantaneous speed at the time of entry (Figs. 1C and 1E). For 20 additional randomly selected bee flights in each of the circular aperture datasets, we extracted the Z coordinate at the X = 10 cm and X = 0 cm positions to examine the approach height (Z) relative to the aperture centres (Fig. S1).

**Motion perception analysis**

The analysis of experimental results involved calculating optic flow as bees flew through circular and stadium-shaped apertures (see retinal velocity heatmap approximating how the optic flow behaves during an aperture traversal in Supplementary Movie 2). Traversals through apertures were captured as single snapshots of position, velocity and optic flow at the moment of aperture entry. This specific time point was chosen because it represents the critical juncture at which bees traverse the aperture. By ensemble-averaging motion data over 30 snapshot samples for each aperture, we mitigated individual variability and preserved the raw dynamics of bee motion at the time of traversing the aperture. While a full trajectory analysis might reveal additional nuances in the approach trajectory, our snapshot approach offers a practical method for inferring potential control mechanisms underlying aperture traversal.

Taking the case of the isotropic circular aperture as an example, we estimated the visual angle $\theta$ subtended by the edges of the aperture between two reference time steps ($t_1$ and $t_2$) to estimate the instantaneous motion of the image of the edge during traversal (Fig. 1E). To quantify optic flow, we calculated the instantaneous translational optic flow ($\Omega$, in $rad\ s^{-1}$) experienced by bees relative to specific edges within the aperture using the simplified formula:

$$\Omega_{t1-t2} = v_{t1-t2}/r_d \qquad (1)$$

where $v$ is the axial velocity (along X) of the observer during traversal, and r is the ray distance to the edge of interest when the bee passes the aperture plane (at $t_2$). The approximation assumes linear and perpendicular motion of the observer relative to the edge, simplifying the calculation at the critical moment of sensory input during traversal. It is derived using the small-angle approximation ($\tan(\theta) \approx \theta$ for small $\theta$), which streamlines the relationship between optic flow and angular motion. While this simplified approach may have limited applicability to more complex, realistic environments, it effectively captures the fundamental optic flow relationship when the sample size is large and enables analyses of the traversal for specific aperture edges.

$$\Omega_{BL} = \frac{\left(\frac{v}{r_{d_{\text{left}}}}\right) + \left(\frac{v}{r_{d_{\text{right}}}}\right)}{2} \qquad (2)$$

$$\Omega_{DV} = \frac{\left(\frac{v}{r_{d_{\text{dorsal}}}}\right) + \left(\frac{v}{r_{d_{\text{ventral}}}}\right)}{2} \qquad (3)$$

$$\Omega_R = \frac{1}{360}\sum_{i=1}^{360} \frac{v}{r_{d_i}} \qquad (4)$$

$$\Omega_V = \frac{v}{r_{\text{ventral}}} \qquad (5)$$

To examine how different aperture regions influence the bee's optic flow during traversal, we computed the optic flow in distinct areas of the visual field (see Eqns 2–5). These include: bilateral optic flow ($\Omega_{BL}$), which sums the motion from the left and right edges (Eqn 2); dorsoventral optic flow ($\Omega_{DV}$), which sums the motion from the top and bottom edges (Eqn 3); radial optic flow ($\Omega_R$), which is the average edge motion calculated by casting rays in 360 directions around the aperture (Eqn 4); and ventral optic flow ($\Omega_V$), which captures the motion from the edge directly beneath the bee (Eqn 5). The unidirectional ($\Omega_V$) and bi-directional optic flows ($\Omega_{BL}$, $\Omega_{DV}$) use rays cast horizontally or vertically from the entry point of the bee to the nearest edge as illustrated in Fig. 3B-E.

**Statistics and data interpretation**

Statistical analyses were performed on the bees' entry location data across apertures of varying sizes, with sample sizes ranging from 100 entries for the smallest aperture to over 500 entries for the larger apertures. The variation in sample sizes was partly attributable to differences in traversal success rates and the necessity to accurately sample the broader spatial distribution inherent in larger apertures. A sample size of 100 entries was deemed sufficient to achieve comprehensive data coverage across the span of the smallest aperture, ensuring that it was adequately represented relative to its absolute area. In contrast, larger apertures required increased sample sizes to effectively capture the expanded spatial distribution, thereby maintaining representative coverage across the larger aperture space.

To facilitate consistent and representative analysis of velocity and optic flow, a stratified random sampling approach was employed. The aperture plane was divided into three equi-spaced spanwise bins on each semi-circular half of the apertures (see vertical dotted lines segmenting the apertures in Fig. 3). From each bin, velocity analysis was performed on the discrete measurements of instantaneous velocity from 10 samples. These samples were chosen from either semi-circular side, resulting in a total of 30 samples (with varying lateral positions) for each optic flow group per aperture before data filters were applied (see Table 1 for sample sizes post-filter for each set). This sampling method minimized central sampling bias by evenly distributing pre-filter dataset sizes across the spanwise bins, ensuring coverage of the entire aperture width.

Throughout this report data are expressed in the form *Median* ± *MAD* to minimise the influence of skewness and outliers, and all motion and position data were ensemble-averaged—aggregating measurements from a large number of samples to provide a representative estimate of bee behaviour. In our visual representations of the raw entry data (Fig. 2), standard deviations are used to collate the raw positional data, facilitating straightforward interpretation and comparison with prior studies that have traditionally employed mean-based metrics.

For OF data, percentile thresholds at the 5$^{th}$ and 95$^{th}$ percentiles were used to exclude the top and bottom 5%, eliminating outliers from measurement errors or atypical behaviours unrepresentative of the general foraging population. Comparing OF variability within and between the datasets was conducted using median-based statistical metrics including the Interquartile Range (IQR), Median Absolute Deviation (MAD) (calculated without scaling), and Coefficient of Variation of the Median ($CV_{Med}$) as shown in Table 1. These metrics are less sensitive to outliers (than mean-based counterparts) and provide a reliable measure of dispersion around the median, making them suitable for skewed data distributions common in behavioural studies. These were considered robust measures of variability in the non-normally distributed data to complement statistical analyses and were used to infer the underlying controllers that the subjects of these experiments might be using.

To assess the statistical significance of differences in optic flow datasets across various aperture sizes and shapes, we employed non-parametric hypothesis testing due to the violations of normality assumptions in the datasets. For pairwise comparisons between two independent groups, we used the Mann-Whitney (MW) U tests. For comparisons involving three or more groups, we used the Kruskal-Wallis (KW) H test to determine whether there were statistically significant differences in the median ranks of independent groups. After obtaining a significant KW test result ($p < 0.05$), we conducted post-hoc pairwise comparisons using Dunn's test with Bonferroni corrections to identify which specific groups differed from each other. We also computed effect sizes to quantify the magnitude of differences. For multi-group comparisons using the KW test, we calculated epsilon-squared ($\epsilon^2$) as a measure of overall effect size for the KW test using $\epsilon^2 = \frac{H-(k-1)}{N-k}$, where $H$ is the KW test statistic, $k$ is the number of groups, and $N$ the total sample size. For pairwise comparisons, we derived a standardized effect size as $r = \frac{Z}{\sqrt{N}}$, where Z is the (signed) z-score corresponding to the MW test result. We reported sample sizes (n), test statistics (H for KW tests, U for MW tests), p-values, and effect sizes (r and $\epsilon^2$) in the results (see Tables S1-3).

## Results

### Bees' flight behaviours in the tunnel

Entry and collision locations of bees were manually digitized to analyse the distributions of aperture traversal locations. Qualitative review of individual flights revealed that bees generally maintained altitude, with vertical movements considerably less pronounced than lateral casting deviations (Supplementary Movies 3 and 4). Before traversing the aperture wall, bees typically aligned themselves with the gap as a preparatory adjustment for imminent passage. Occasionally, bees approached at non-orthogonal angles—typically near the sides—and exhibited increased lateral motion. Overall, the bees' approach flight path was relatively well aligned with the central axis of the tunnel from the inlet to the feeder, suggesting that the modest vertical adjustments serve both to minimise the risk of edge collisions and to optimise the perception of aperture geometry. At 10 cm upstream from the transverse wall (X = 10 cm), for all circular aperture diameters (30 - 150 mm), the Z-coordinate of bees relative to that at the aperture plane (X = 0 cm) was non-significant, suggesting that bees had a nominally straight and level approach (Fig. S1 and Table S1). Following the initial approach, bees traversed the apertures at varying speeds and positions relative to the aperture before proceeding towards the feeder platform on the opposite side of the wall.

Bees that failed to avoid collisions during passage through the aperture typically displayed a poorly controlled trajectory immediately after the traversal point. In cases of wing collisions, bees experienced a subsequent roll toward the impacted side and a drop in altitude, often followed by an overcorrection that resulted in rapid lateral displacement and oscillatory roll adjustments en route to the feeder. Body collisions were predominantly observed when bees executed lateral manoeuvres near the aperture edges, while leg collisions were less pronounced, generally involving only superficial contact with the bottom edge of the aperture (Supplementary Movie 4).

### Entry and collision locations during aperture traversals

*Circular apertures*

Tests were divided into three categories: (A–E) circular apertures with varying diameters (30 mm to 150 mm), (F–G) mid-sized apertures (60 mm diameter) with extended lateral or vertical sides (created by adding straight-edged sections between two semi-circular halves), and (H–K) mid-sized circular apertures (60 mm diameter) with varying surrounding visual textures (Fig. 3 and Table 1). For each aperture type, we digitized the vertical (Z-axis) and lateral (Y-axis) entry coordinates of at least 100 bee flights, as shown in Fig. 2A-K, providing a detailed view of the distribution of bee entry positions across aperture geometry types.

Across all aperture types, the median lateral entry position was centered between the left and right edges (Table 1, Fig. 2A-K), indicating that bees tended to enter at the bilateral midpoint of the aperture. In contrast, the vertical entry position (Z-axis) showed a consistent tendency for bees to pass through the lower half of the aperture—a trend that became more pronounced with larger circular apertures. For the smallest circular aperture (A, D = 30 mm), most bees entered near the center of the aperture (median lateral position ($\tilde{Y}$ = 0.21 mm), median vertical position ($\tilde{Z}$ = 0.24 mm), as indicated by the blue median marker in Fig. 3A. In the largest aperture (E, D = 150 mm), the median vertical entry position was significantly below the center ($\tilde{Z}$ = −24.9 mm), while the median lateral position remained approximately centered ($\tilde{Y}$ = 1.06 mm).

Smaller apertures exhibited lower positional variation as shown by the standard deviations (and mean absolute deviations) in both lateral and vertical entry positions (see green boxes in Fig. 3). For the smallest aperture, the standard deviations were $\sigma_Y$ = 6.56 mm (lateral) and $\sigma_Z$ = 5.51 mm (vertical). These values increased approximately linearly with aperture size, reaching $\sigma_Y$ = 22.44 mm and $\sigma_Z$ = 29.28 mm for the largest aperture (Fig. 2A-E). This trend persisted in proportion to the aperture radius across all sizes of the tested circular apertures (Table 1).

Collisions with the edges of the circular apertures were more frequent in smaller apertures compared to larger ones (Fig. 2A-E and Table 1). In the smallest aperture (A, diameter 30 mm), 61.5% of the bees experienced collisions, whereas only 10.1% did so in the largest aperture (E, diameter 150 mm). In aperture A, most collisions (96.7%) involved the bees' wings, with less than 10% involving body collisions upon entry or during traversal (Table 1). Bees that made wing contact with the rim experienced significant roll perturbations post-traversal, adversely affecting their flight stability and often resulting in downward veering to one side. Bees that made edge contact involving their bodies (abdomen, thorax, or legs) were less impacted in their subsequent flight behaviour (Supplementary Movie 4).

In general, smaller circular apertures showed a higher proportion of wing collisions compared to body collisions (Fig. 2A-E and Table 1). In larger apertures, most of the fewer observed collisions were body strikes, comprising 82.3% of collisions in aperture E (Table 1). Wing collisions generally occurred in the upper half of the apertures, while body collisions were more common in the lower half (Fig. 2). It was common to observe bees with body strikes 'brushing' their legs against the bottom edge of the aperture. Although these instances were recorded as collisions, they had minimal influence on the bees' flight trajectories. Pre-traversal flight was qualitatively indistinguishable between collision statuses, suggesting that there is a broad behavioural buffer between what we classify as collision and collision-free flights. Given that the ratio of collision flights is small for most aperture sizes, we focused our speed analysis on collision-free samples to reliably capture the underlying behavioural strategies. A dedicated analysis of collision flights, which could further elucidate differences in behaviour, remains a promising direction for future work.

*Apertures with modified shape*

Flights through the horizontally and vertically extended stadium-shaped apertures (Fig. 2F–G) demonstrated the influence of aperture shape on bee entry locations. These apertures were created by retaining the curvature of the 60 mm circular aperture while extending only the width (horizontal elongation, aperture F) or the height (vertical elongation, aperture G). In the horizontally elongated aperture, bees traversed at a position closer to the lower edge relative to the aperture height (median vertical position $\tilde{Z} = -9.88 \pm 9.25$ mm, h = 60 mm), whereas in the vertically elongated aperture, bees traversed relatively closer to the center ($\tilde{Z} = -12.31 \pm 19.39$ mm, h = 120 mm). In both cases, bees maintained horizontal centering (median lateral positions $\tilde{Y} = 2.21 \pm 15.27$ mm and $\tilde{Y} = -1.20 \pm 12.13$ mm, respectively). The mean absolute deviation was greater in the direction of elongation for each aperture, and relatively greater for the vertically elongated shape ($MAD_Y = 15.27$ mm, $MAD_Z = 9.25$ for aperture F, and $MAD_Y = 12.13$, $MAD_Z = 19.39$ for aperture G, Table 1).

Only 10% of bees exhibited edge collisions during traversal of the horizontally elongated aperture (F), approximately three times less than the 60 mm circular aperture (28%, Table 1). More than 50% of bees experienced collisions in the vertically elongated aperture (Fig. 3 and Table 1 F-G), which was more than double the rate of collisions in the 60 mm circular aperture. Of these, 67% were body/leg collisions in the horizontal aperture, with more than half brushing their legs or touching the lower edge. In the vertical aperture, 94% of all collisions were wing collisions, comparable to the ratio for the 60 mm circular aperture (Table 1). Body collisions in both elongated shapes tended to occur at the lower sections of the aperture rim similar to all other tested apertures as shown in Fig. 3F-G.

In the horizontally elongated aperture, bees' entry locations were proportioned similar (both close to the -1/3 normalised height) to those in the circular aperture of the same lateral major-axis scale (120 mm, Aperture D) as seen in Fig. 2F and Fig. 2D. However, due to aspect ratio differences, absolute entry heights were reduced compared to both the 60 mm and 120 mm circular apertures (Fig. 2F and Table 1). Being closer to the edge below, they traversed at significantly slower speeds than the 120 mm aperture (corresponding to the major axis in width) and similar speeds as the 60 mm aperture (corresponding to the minor axis in height of the horizontally elongated aperture). In the vertically elongated aperture, bees also flew at entry locations proportional to the 120 mm aperture (in height) but at significantly lower speeds slower that were comparable to both the 60 mm and horizontally elongated apertures (Fig. 4A). However, they flew at absolute heights significantly greater than both the 60 mm and 120 mm diameter circular apertures. This resulted in significantly lower ventral optic flow compared to all other apertures (Fig. 4B-E and Table S2).

*Apertures with modified surrounding texture*

To investigate the role of surrounding textures on traversal, we examined the effect of removing visual textures surrounding specific halves of the 60 mm circular aperture: lower, upper, right, or left (Fig. 2H-K). The distributions of entry locations were biased towards the texture-flanked half as shown in Table 3. The bias was significant when comparing the group bilateral entry distributions for left and right (J vs K) texture-flanked halves with the 60 mm plain texture aperture (B), as indicated by the significant p-value yielded from the omnibus test of three groups ($p < 0.05$) (Table 3). Subsequent pairwise tests between the three groups revealed a significant comparison only between the left flanked and the right flanked datasets ($p = 0.0036$), but not against the uniform control (Table 3). Biases were insignificant when group-comparing the horizontal and vertical coordinate datasets of upper and lower texture-flanked halves and the uniform control ($p = 0.750$ for Y, $p = 0.103$ for Z coordinates). Although the overall group comparisons test did not yield significance, Dunn's post-hoc pairwise tests were performed for all pairs to maintain consistency with the comparison of side-flanked set where a significant test result was observed (see linked data repository). The pairwise comparison test yielded a p-value of 0.1004 for the Z coordinates between the two vertically flanked tests (H and I), which approaches the significance level but does not meet the criteria for confirming the trend.

Collisions for each texture-flank location and aperture edge are shown in the linked data repository. An omnibus Chi-Square Test for Homogeneity indicated no overall difference in distribution of collisions across edges among the five edges when flanked with different locations of texture ($\chi^2 = 11.95$, df = 12, $p = 0.449$). To probe potential differences between collisions on specific sides, we performed separate 2×4 chi-square tests (unadjusted). Only the comparison between "Left Flanked" and "Right Flanked" reached nominal significance ($p = 0.047$), but this difference does not remain significant after a Bonferroni multiple-comparison correction. We therefore conclude that there is no statistically meaningful difference between any pair of collision-texture location conditions once we control the familywise error rate.

**Visual motion results across aperture types**

The traversal velocity of bees increased with the diameter of the circular apertures (Fig. 3A). In the largest aperture (E, 150 mm), bees' median flight traversal speeds ($v_x = 0.57 \pm 0.15$ m s$^{-1}$) approached the speeds in the tunnel without any aperture wall obstruction ($v_x = 0.76 \pm 0.15$ m s$^{-1}$). The total optic flow produced by the edges of the apertures generally increased with aperture diameter, peaking for apertures between 60 mm and 90 mm diameter and then decreasing for larger apertures across $\Omega_{BL}$, $\Omega_{DV}$, and $\Omega_R$ OF measurements. For circular apertures, the median ventral optic flow magnitudes ($\Omega_V$) ranged narrowly about the median of 767 ± 174 deg s$^{-1}$ (IQR = 269 deg s$^{-1}$), as shown in Fig. 3B-E., exhibiting less absolute variation than in the other OF measurements (Tables 2 and S2).

The ventral OF metric ($\Omega_V$), derived from a single ray, showed a $CV_{Med}$ of 0.227 across circular apertures of varying diameter (30-150 mm), which is lower or comparable to the values obtained for multi-ray metrics for the same apertures: bilateral (two rays, $CV_{Med} = 0.292$), dorsoventral (two rays, $CV_{Med} = 0.237$) and radial (360 rays, $CV_{Med} = 0.250$) (Fig. 3B-E and Table 2). Theoretically, averaging across multiple rays should reduce variability by approximately a factor of $\sqrt{2}$ for two-ray configurations, with even larger reductions expected for radial measurements. The multiple-group statistical comparisons revealed that the ventral OF had no significant differences across the circular aperture diameters ($p = 0.124$), whereas the other multi-ray metrics displayed numerous significant differences ($p < 0.001$) (Tables 3 and S2-3). The relatively narrower variability in the ventral OF datasets is indicative of an intrinsic stability in this metric compared to the others.

Velocities through the horizontally and vertically elongated apertures (F-G) were not statistically different than those through the circular aperture of the same minor axis diameter (Aperture B, 60 mm diameter) (Fig. 4A). The speeds and bilateral and dorsoventral optic flow measurements compared between apertures F and G were significant (Fig. 4A-C), suggesting strong influence of aspect ratio on the traversal strategies of bees.

Bees displayed similar speeds when flying through the reduced-texture 60 mm apertures (H-K, Fig. 5A), with a median velocity of $v_x = 0.31 \pm 0.13$ m $s^{-1}$, compared to the fully textured 60 mm aperture ($v_x = 0.30 \pm 0.09$ m $s^{-1}$ (Fig. 5A). The non-parametric group comparison test (KW) returned a non-significant p-value ($p = 0.889$), indicating no significant difference in traversal speed between the different aperture conditions. The optic flow measurements across all apertures showed similar relative variabilities and were not significantly different when compared among each other (Fig. 5B-E, Tables 2 and S2-3).

## Discussion

### Centralized traversal of narrow apertures

Our study investigated how bees navigate through apertures of varying sizes and shapes, challenging the hypothesis that they would consistently fly through the center—a path theoretically offering maximal clearance and minimal collision risk. When navigating through small apertures (30-60 mm diameter), bees exhibited geometric centering behavior both laterally and vertically, minimizing the risk of collision with the edges (Fig. 3). This centering behavior is consistent with the idea that bees rely on the balance of symmetrical visual cues from the surrounding edges to guide their flight in confined spaces (Srinivasan et al., 1996; Srinivasan and Zhang, 1997). The proximity of both lateral and vertical boundaries provides strong bi-directional optic flow cues that enable bees to maintain a centered trajectory.

Analysis of lateral positions across all aperture diameters showed that the absolute standard deviations of bee entry positions increased proportionally with the widths of the apertures (Table 1). However, relatively, lateral entry positions of bees were consistently confined within the central one-third band of the apertures, regardless of diameter (Fig. 2). This pattern was maintained in the elongated apertures, indicating that bees regulate their lateral positioning relative to the proportions of the aperture. Tighter absolute spatial constraints in smaller apertures demand flight precision within the narrower absolute dimension of the band. By maintaining a proportional distance within the physical boundaries of the aperture, bees seem to optimize their movement for safest passage with respect to their wingspan.

As the aperture size increased, bees displayed less vertical centering behavior, opting instead to fly closer to the lower edge of the aperture. This shift suggests that in larger apertures, where spatial constraints are reduced and the risk of collision is lower, bees do not prioritize centering; instead, they adopt alternative strategies, possibly relying more on ground-based visual cues. The decreased vertical centering in larger apertures and the tendency toward the lower edge imply that bees are using a flight strategy based on a preferred fixed edge: the bottom edge. In large spaces, where the consistency in relative standard deviation represents a greater absolute dimension (vertical lines of the green boxes in Fig. 2), the need for precise centering diminishes.

The transition from centering behavior in small apertures to favoring flight near the lower rim in larger apertures may reflect an inherent trade-off between the benefits of centering and the costs associated with maintaining it. Precise centering requires continuous processing of visual cues from both bilateral and dorsoventral fields and frequent course corrections, which may be energetically costly. In larger apertures, the reduced necessity for such lateral precision could allow bees to simplify their flight control and conserve energy for other complex behavioral demands.

**Adaptive traversal altitude**

Building upon our observations that bees center their flight relative to aperture edge curvature, we investigated the influence of the ventral edge on entry height distributions. We grouped data from pairs of bins equidistant from the center (Outer: bins 1 and 6; Middle: bins 2 and 5; Inner: bins 3 and 4) to represent regions of varying curvature along the ventral edge, forming vertical trisectors (Fig. 6A-B).

After normalizing entry heights by aperture radius (R), we found that in the outer trisector where vertical space is most constrained (1.1 R), bees consistently maintained an entry height of approximately 0.5-0.6 R above the ventral edge across all aperture sizes (Fig. 7B). The bees aligned near the vertical center of the available space when traversing the outer third of circular apertures, consistent with the vertical centering behaviour observed in small diameter-constrained apertures (Fig. 3A-E). In the inner bins near the aperture center (where vertical space is less constrained) the bees' normalized entry heights varied more with aperture size (Figs. 2A-E and 6B). In the smallest apertures, bees in the innermost bins exhibited higher median normalized heights of around 1R above the ventral edge (Fig. 6B), effectively maximizing clearance.

The bees' tendency to maintain this relative height when free of spatial pressures suggests reliance on a simple proportional metric (like our division into proportional sectors) as a simple strategy for safe traversal, requiring minimal computational effort by leveraging the aperture's relative geometry during flight. By maintaining an entry height of 0.5-0.6 R when space is imposing in a particular vertical segment, collision avoidance is optimised in varying aperture geometries, particularly within the outer trisector. Considering additional factors, the abrupt change in optic flow experienced during the transition from the visually constrained aperture to open tunnel space (immediately after traversal as shown in Supplementary Movie 2) could necessitate rapid adjustment. This may underlie the observed reduction in altitude post-traversal, as bees attempt to locate the edge that they have just passed.

### Ventral optic flow controller in a narrow visual field

Speed is proportional to the aperture entry location relative to the edges, as bees fly slower near surfaces which produce large OF signals (Srinivasan et al., 1996). Incorporating speed into our analysis of flight altitude reveals that the observed vertical distances to the edge below are correlated to ventral OF, with bees maintaining ventral edge distances that on average correspond with their flight speed, thereby keeping ventral OF in a consistent range (see Fig. 7). In larger apertures, bees entering near the sides (where the vertical distance to the edge below is lower due to rim curvature) maintained velocities comparable to central regions by raising their flight altitude (no significant velocity differences between trisectors in Fig. 7A). This altitude adjustment allows them to preserve consistent ventral OF without significant changes in speed (Fig. 7B). However, in smaller apertures, the limited space between the dorsal and ventral edges restricts altitude adjustments, creating a dilemma that leaves centering as the only viable option. As a result, bees are forced to reduce their speed near the edges to keep ventral OF consistent.

One prevailing hypothesis is that bees process optic flow across their entire visual field to detect discontinuities and locate the center of an aperture. However, this strategy might be complex and error-prone due to the difficulty of integrating extensive visual information. Frontal OF (and associated time-to-contact cues) could assist in detecting features along the direction of flight; however, in most trajectories the bees were not on a collision course (as shown by the low collision ratios in Table 1, and average entry positions distant from edges in Fig. 2). Consequently, its value is lower than the steadier single-edge feedback, consistent with the total OF being less stable than edge-specific OF (Fig. 3). An alternative, simplified flight control strategy that focuses on stabilizing ventral OF, which is key data in other modes of flight, could be more effective. This approach would enable bees to navigate safely through narrow, elevated passages by intuitively measuring their speed and distance relative to proximal edges beneath the observer.

Across circular apertures of all sizes and at all lateral locations of entry, bees maintain a ventral OF with a relatively confined range of approximately 767 deg $s^{-1}$ (Fig. 7B). In larger apertures, flying near the side edges reduces the vertical distance to the ventral edge, causing a surge in the perceived passing rate of edges in the scene beneath, thereby increasing ventral OF. In response, a ventral OF controller could adjust elevation accordingly to optimise clearance. In smaller apertures, where dorsoventral balancing pressure may oppose altitude increases, ventral optic flow can be sustained by reducing velocity.

### Role of ventral optic flow cues in elongated apertures

We evaluated the effectiveness of the proposed ventral OF controller in elongated apertures (apertures F and G). If bees utilize a simple OF controller based on their height above the lower edge and speed adjustments, we would expect similar behavioural patterns relative to the OF generated by the lower edge in both circular and non-circular apertures. However, our observations revealed deviations from this expectation (see Fig. 4E). In horizontally elongated apertures, bees flew at lower altitudes and maintained similar speeds compared to the 60 mm circular aperture. In vertically elongated apertures, bees flew at higher altitudes above the bottom edges and at slower speeds compared to other relevant apertures, resulting in a significantly lower ventral OF.

These findings indicate that an aperture's aspect ratio influences the relative weighting of OF cues for speed regulation and positioning. The results imply that bees exhibit a behavioural response to the shape of non-circular apertures, potentially perceiving their spatial layout in relation to their own body dimensions and adjusting their flight paths accordingly to minimize collisions, as observed in other gap challenge examples in the literature (Ravi et al., 2020). In narrow vertical apertures, reduced horizontal width increases lateral OF from closer side edges. This could prompt bees to maintain a more central position, potentially compensating for the weaker ventral OF caused by the more distant lower edge. This behaviour suggests that bees rely more heavily on lateral OF cues when they are more prominent, similar to how they regulate speed when flying through narrow corridors with close side walls (Srinivasan et al., 1996; Baird et al., 2010).

**Influence of visual texture on aperture passing strategies**

The preferential entry toward texture-flanked sides seen in the results suggests that bees utilize visual textures as key cues for aperture traversal. They favor areas with richer visual information, tending to enter closer to those sides, with a stronger preference towards laterally arranged textures compared to vertically arranged ones. Anatomically, bees' have higher resolution in the frontal regions due to the ommatidia arrangement in their compound eyes, which augments their sensitivity to lateral and ventral cues (Hecht and Wolf 1929; Seidl and Kaiser 1981). Additionally, their flight dynamics, characterized by lateral motion inertial sensitivities and thrust vectoring for control (Ravi et al., 2013; Crall et al., 2015), facilitate rapid horizontal repositioning during traversal, which could explain why there was stronger bias toward lateral textures.

Despite the biases, traversal speeds remained consistent across all aperture texture conditions (Fig. 5A, $p = 0.889$), indicating that reduced surrounding texture has little influence on traversal OF cues. Similarly, the absence of significant differences in optic flow measurements across texture conditions (Fig. 5B-E) supports the notion that bees primarily rely on the detection of edges, rather than the surrounding surface texture, for shape discrimination and navigation. This aligns with previous literature indicating that bees use contrast and brightness gradients in their photoreceptor channels, making untextured edges less discernible than textured ones (Lehrer et al., 1990; Horridge, 2015; Srinivasan and Lehrer, 1988). The contrasting may allow them to perceive aperture boundaries upstream, gleaning geometric data on the obstacle before reaching the target challenge. The bias toward textured sides may be due to the enhanced contrast and visual salience provided by textures in particular regions during approach, facilitating preparatory localisation before transit.

Thus, while richer visual textures may enhance the visual salience of certain sides of the aperture, facilitating earlier and more confident localization, the fundamental cue for controlled passage appears to be the presence and orientation of edges. Once edges are detected, bees could rely on OF generated by these boundaries—especially the ventral edge—to guide their movement. This combination of textural preference during approach and edge-based navigation during the fine control needs of traversal underscores a sophisticated and adaptive visual guidance strategy in bees.

### Implications for aerial navigation in insects

Apertures are ubiquitous in both natural and urban environments, making the ability to efficiently perceive and navigate through them indispensable for flyers like bees and other insects. Our findings suggest bees may prioritizes certain optic flow cues, particularly from the ventral edge of an aperture, when adjusting trajectories through constricted spaces. Although our findings do not definitively confirm the existence of a dedicated ventral optic flow controller, this selective use of visual field regions aligns with theories tying localized visual processing tactics to specific flight tasks (Lecoeur et al., 2019). Such strategies in monocular vision-based flyers highlight optimised usage of visually derived signals in many flight contexts. These insights deepen our understanding of insect navigation strategies and can serve as inspiration for designing bio-inspired, resource-efficient algorithms for robotic navigation in complex environments.

### Limitations and areas for future work

In this study, we simplified the setup and analysis, leveraging large samples sizes to reduce natural variance and draw generalisations. The simplified analysis revealed general behavioral tendencies that reflect the underlying control mechanisms. A key simplification was that our optic flow analysis focussed on translational optic flow and assumed linear translation perpendicular to the aperture edges. The findings should be interpreted as evidence consistent with the postulated control mechanisms rather than a comprehensive account of the underlying sensorimotor processes.

Another potential limitation is that the experimental design relied on bees from a single colony, thereby limiting genetic diversity. Though several dozens of individual foragers were noted individually, we also did not identify individual bees, which raises the possibility of pseudo-replication, although the high frequency of foraging flights and our training protocol served to minimise repeated sampling from the individuals. Additionally, while the bees were trained using standard food-following methods consistent with established procedures in similar studies, the potential influence of training on navigational behaviour cannot be entirely ruled out.

A detailed investigation into discrete bee trajectory manoeuvres during aperture traversal, which includes lateral scanning and comprehensive analysis of approach flight dynamics and differentiating colliding and non-colliding bees, would help elucidate the definitive mechanisms by which honeybees perceive aperture size and determine traversal flight paths.

**Competing Interests**

No competing interests declared.

**Funding**

This research was partially supported by the Australian Government through the Australian Research Council's Discovery Projects funding scheme (project DP220101883) and by the Asian Office of Aerospace Research and Development (AOARD) (Grant No. FA2386-21-1-4075 and FA2386-23-1-4033).

**Data and Resource Availability**

All relevant data and resources are available within the article and its supplementary information.

**Figure Legends**

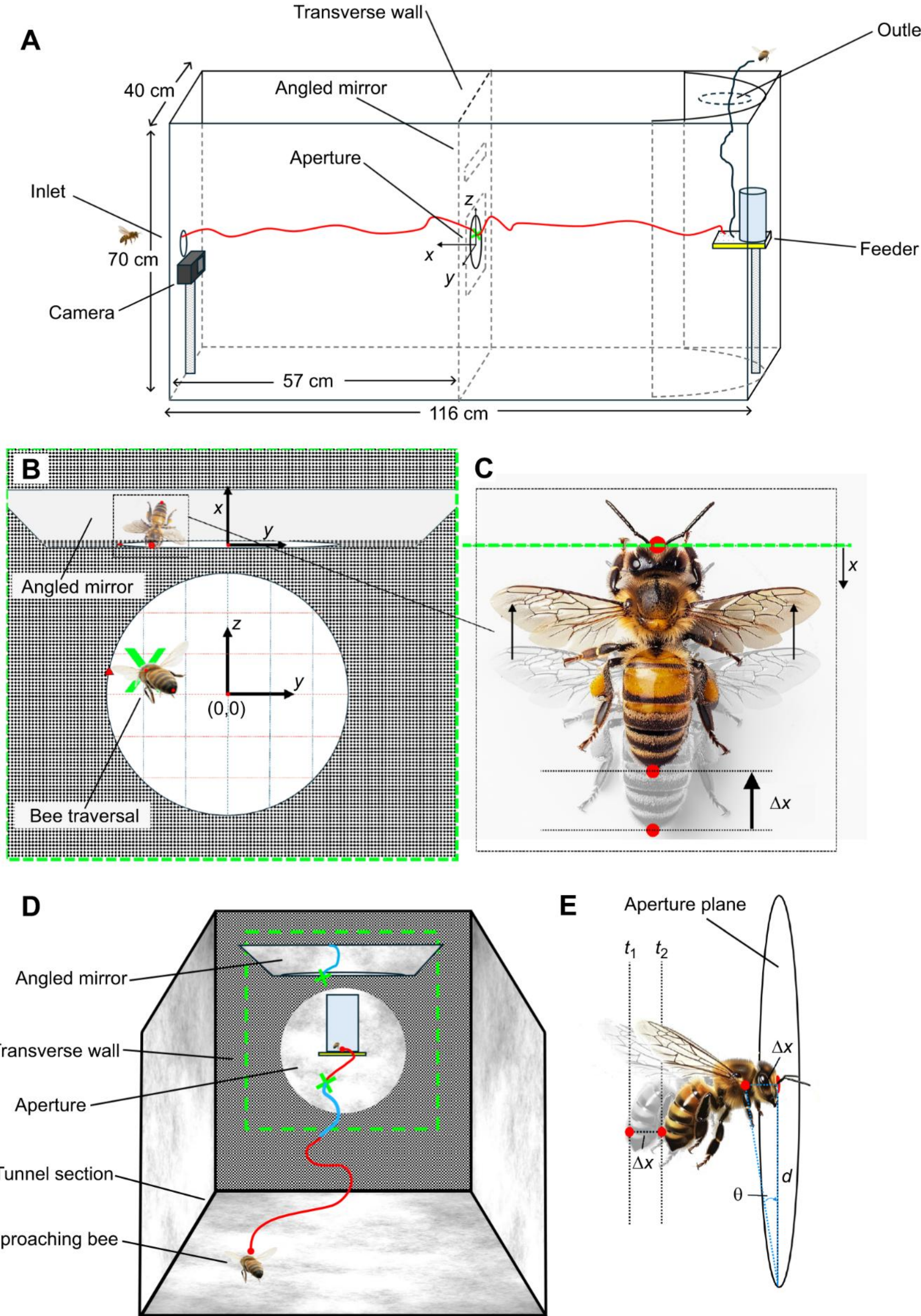

**Fig. 1. Schematic representation of the tunnel configuration, illustrating the experimental procedure and the frames of reference used for digitising traversal position and velocity.**

(A) Perspective view showing the typical flight path of honeybees (red lines) as they entered through the inlet, flew along the tunnel, and passed through an interchangeable aperture located centrally in the transverse wall. The aperture provided access to a separate feeding chamber containing a sucrose solution feeder positioned on a platform at the far end. Following feeding, bees exited via an overhead outlet. The coordinate axes (X, Y, Z) indicate the frame of reference used for analysis.

(B) Front view of the transverse wall and aperture from the camera's perspective, showing the coordinate axes, checkerboard pattern, and the angled mirror. The angled mirror, mounted above the aperture, provides a simultaneous view of the XY plane. The large green cross is shown at the location of an example entry point, and a red triangular marker marks an example wing collision. The distil tips of the bee, used for speed analysis, are highlighted by red dots.

(C) Plan view (XY) of the bee traversal as seen through the angled mirror. The black arrows indicate the X and Y axes, while the origin (0,0) and an example point are marked in red. The aperture plane is indicated by a green line, orthogonal to the X axis. The coloured and greyed-out bee illustrations represent the position of the bee at two distinct time points ($t_1$ and $t_2$).

(D) Perspective view of the experimental setup as recorded by the camera aligned with the principal axis of the tunnel leading to the feeder at the far end. A representative bee trajectory is depicted by a red line, with the segment captured by the angled mirror highlighted in blue. The dashed rectangular outline indicates the aperture plane, corresponding to the close-up view displayed in (B).

(E) Schematic side view illustrating how the optic flow was derived from the visual angle subtended by the edges of the aperture on the bee's retina. In this example, the calculation focusses on the bottom edge of the aperture. The time t1 corresponds to the frame before the bee crosses the YZ plane, and t2 is the frame at which the bee begins to traverse the aperture. The greyed-out bee depicts the t1 position, while the coloured bee shows the t2 position. The rear tip of the abdomen (red dot) was tracked at both time points (using the mirror view) to estimate the forward speed.

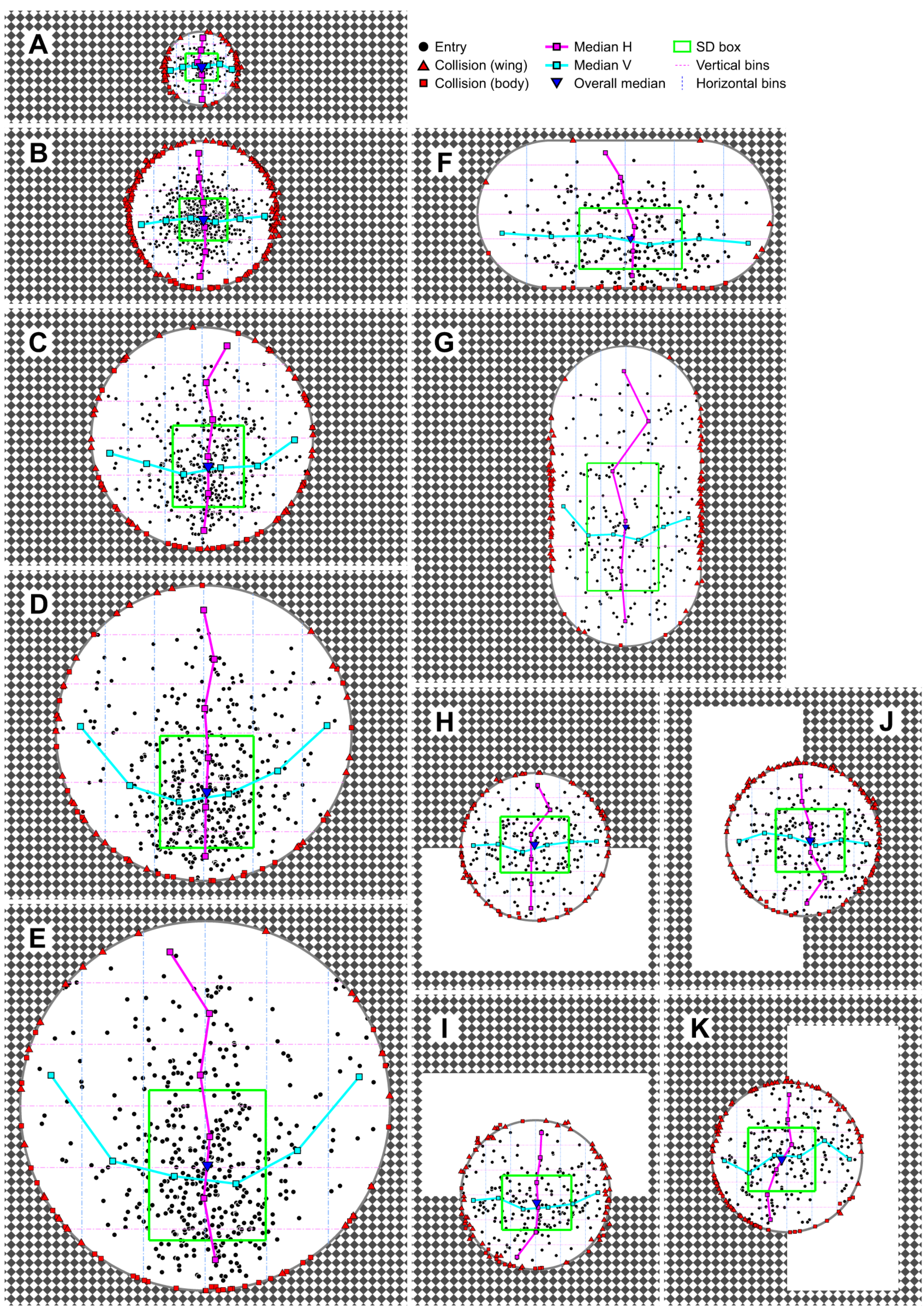

**Fig. 2. Composite panel of bee entry and collision locations for all tested apertures**.

(A-K) Apertures are categorised into three types: circular (A-E), stadium (F-G), and side-flanked texture (H-K). Each aperture is enclosed within a white box and labelled with its corresponding tile letter. Black markers denote entry points, while red markers (triangular or square) indicate collisions (wing or body, respectively). The SD boxes (green lines) depict the area encompassing ±1 standard deviations from the median bee horizontal (Y) and vertical (Z) entry positions. Six equidistant bins are defined along the horizontal (blue) and vertical (magenta) axes, delineated by dotted lines. Within each bin, square markers (magenta for horizontal bins and cyan for vertical bins) represent the median entry locations; thick lines connect these medians across bins for each axis. A single blue triangular marker in each aperture denotes the overall median entry position. To enhance figure visibility and reduce visual strain, the background pattern was rendered in a dimmed contrast tone relative to the figure. Note that in the experiments, black diamonds were used on a white background. Aperture shapes are scaled relative to the actual dimensions of the background pattern. For apertures with altered surrounding texture (H-K), the blank half of the transverse wall is shown as a small white rectangle, reflecting the configuration in which the entire half-wall region on the corresponding side (North, South, East, or West) was left unpatterned.

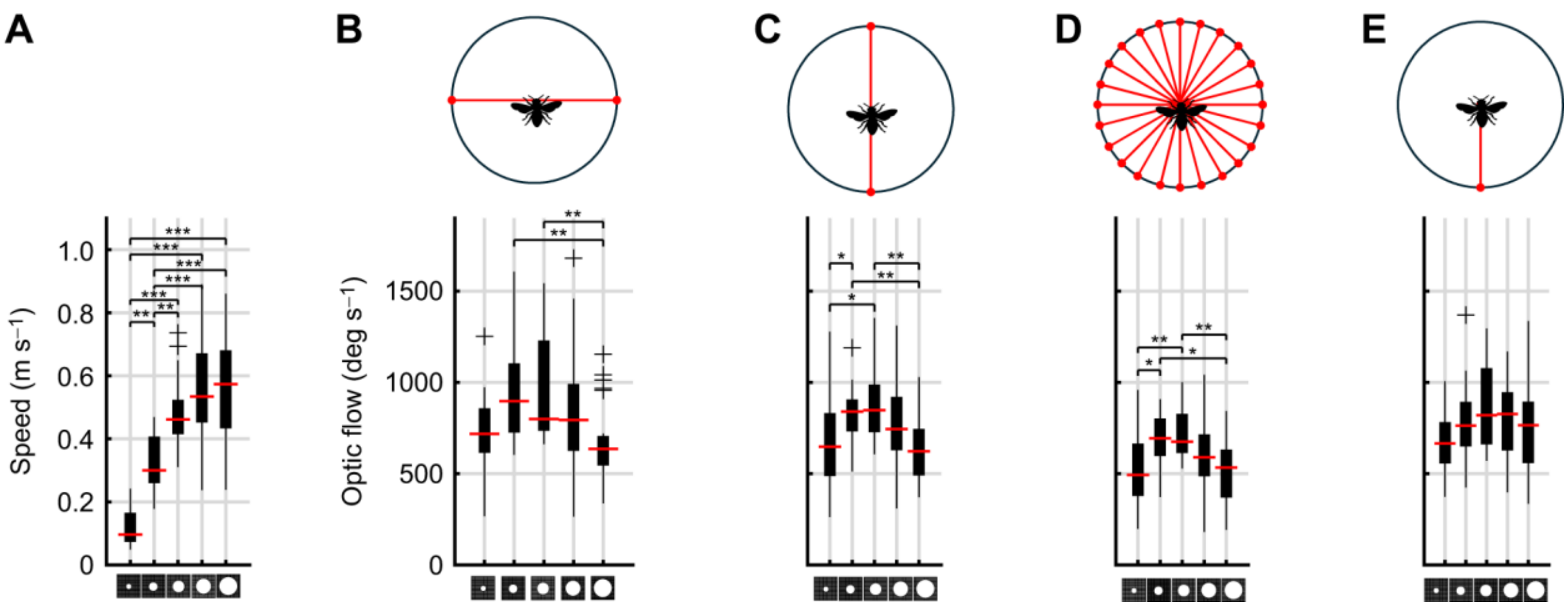

**Fig. 3. Bee traversal speed and optic flow for circular apertures with uniform surrounding texture.**

(A) Traversal speed measured along the longitudinal (X) axis of the tunnel for apertures ranging from 30 to 150 mm in diameter.

(B-E) Four optic flow metrics for these same apertures: bilateral optic flow ($\Omega_{BL}$), dorsoventral optic flow ($\Omega_{DV}$), total radial optic flow ($\Omega_{R}$), and ventral optic flow ($\Omega_{V}$). Each boxplot shows the median in red, plus signs (+) as outliers, and significant differences indicated by asterisks (* for $p < 0.05$, ** for $p < 0.01$, and *** for $p < 0.001$). Diagrams above each sub-panel depict the reference axes (rays) used to estimate the visual motion angles subtended by the corresponding edges. Icons beneath each axis represent the aperture type. Sample sizes (n) and variability statistics for each measurement are provided in Table 2.

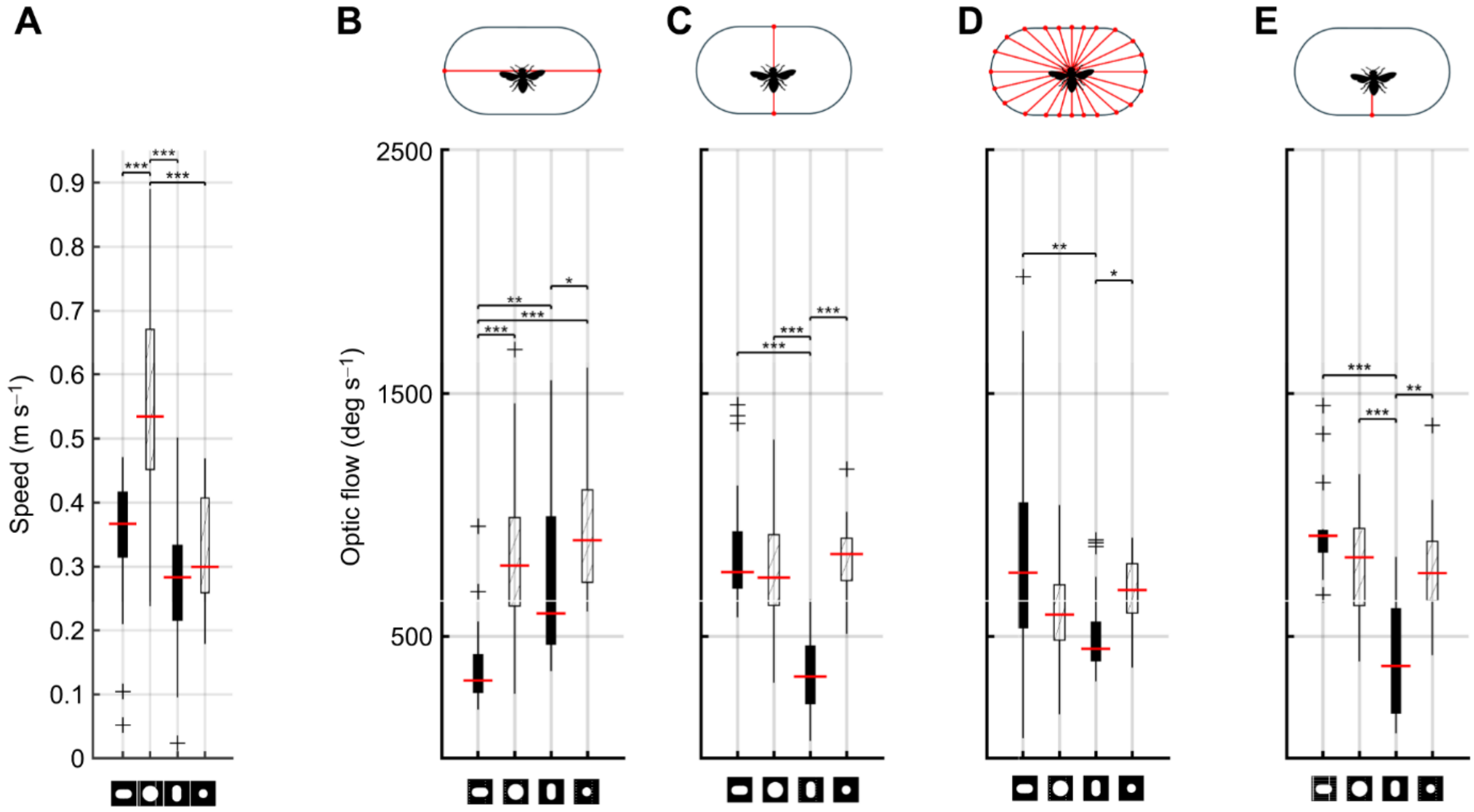

**Fig. 4. Bee traversal speed and optic flow for elongated (stadium-shaped) apertures.**

(A) Traversal speed measured along the longitudinal (X) axis of the tunnel for the two elongated apertures.

(B-E) Four optic flow metrics for these apertures: bilateral optic flow ($\Omega_{BL}$), dorsoventral optic flow ($\Omega_{DV}$), radial optic flow ($\Omega_{R}$) and ventral optic flow ($\Omega_{V}$). Each boxplot displays the median (red line), interquartile range, and potential outliers (+). Asterisks indicate significant differences between groups (* for < 0.05, ** for p < 0.01, *** for p < 0.001). Diagrams above each optic flow sub-panel depict the reference axes (rays) used to estimate the visual motion angles subtended by the corresponding edges. Icons beneath each axis represent the aperture type. Full statistics for each measurement are provided in Table 2.

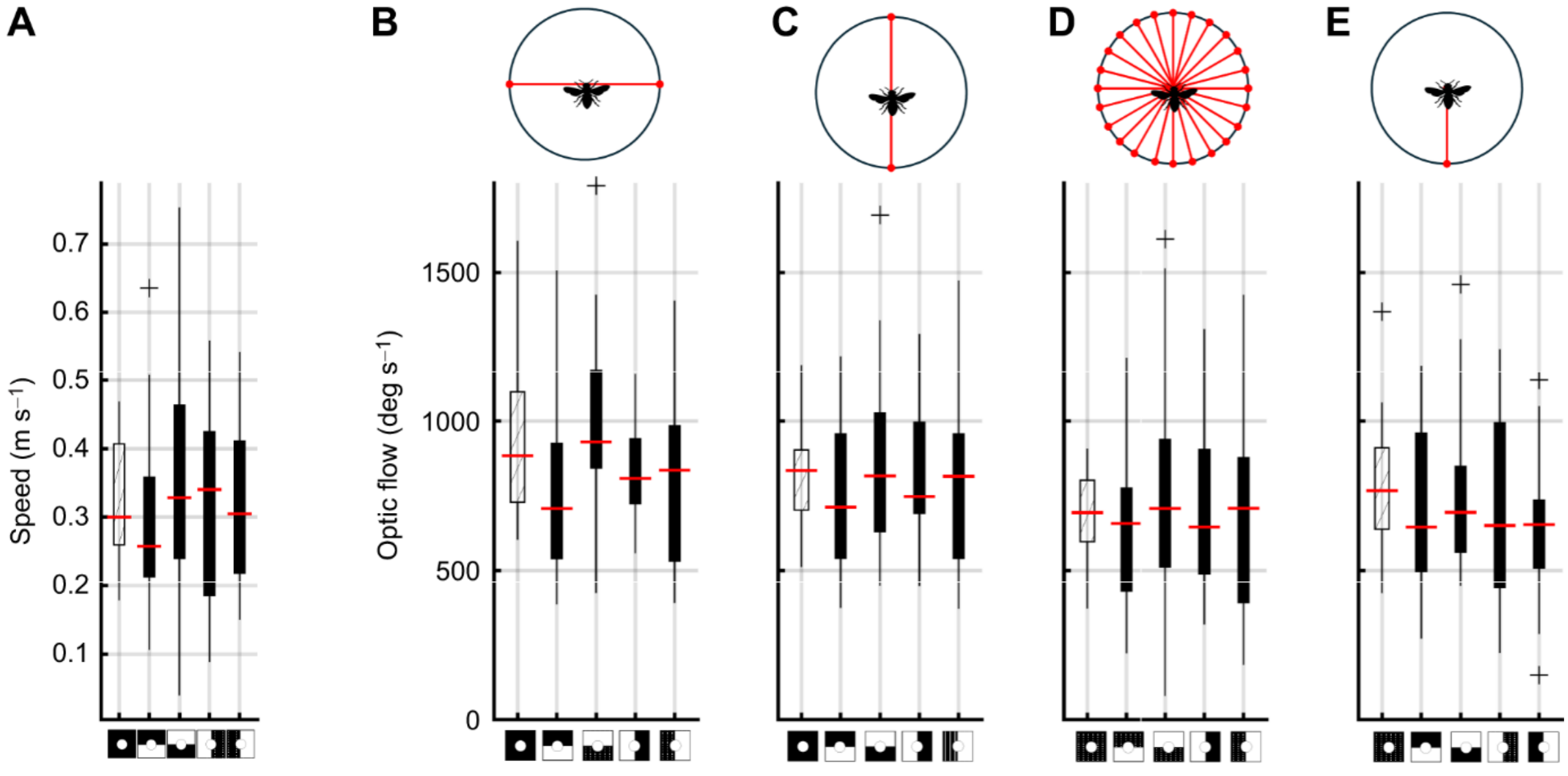

**Fig. 5. Bee traversal speed and optic flow for 60 mm apertures with varying surrounding texture (North, South, East or West half-flanked).**

(A) Speed measured along the longitudinal (X) axis of the tunnel setup.

(B-E) Four optic flow metrics for these partially textured apertures: bilateral optic flow ($\Omega_{BL}$), dorsoventral optic flow ($\Omega_{DV}$), radial optic flow ($\Omega_R$) and ventral optic flow ($\Omega_V$). Data from the fully patterned 60 mm aperture test are included for comparison. Diagrams above each optic flow sub-panel depict the reference axes (rays) used to estimate the visual motion angles subtended by the corresponding edges. Icons beneath each axis represent the aperture type. Full statistics for each measurement are provided in Table 2. No significant differences were detected among the groups.

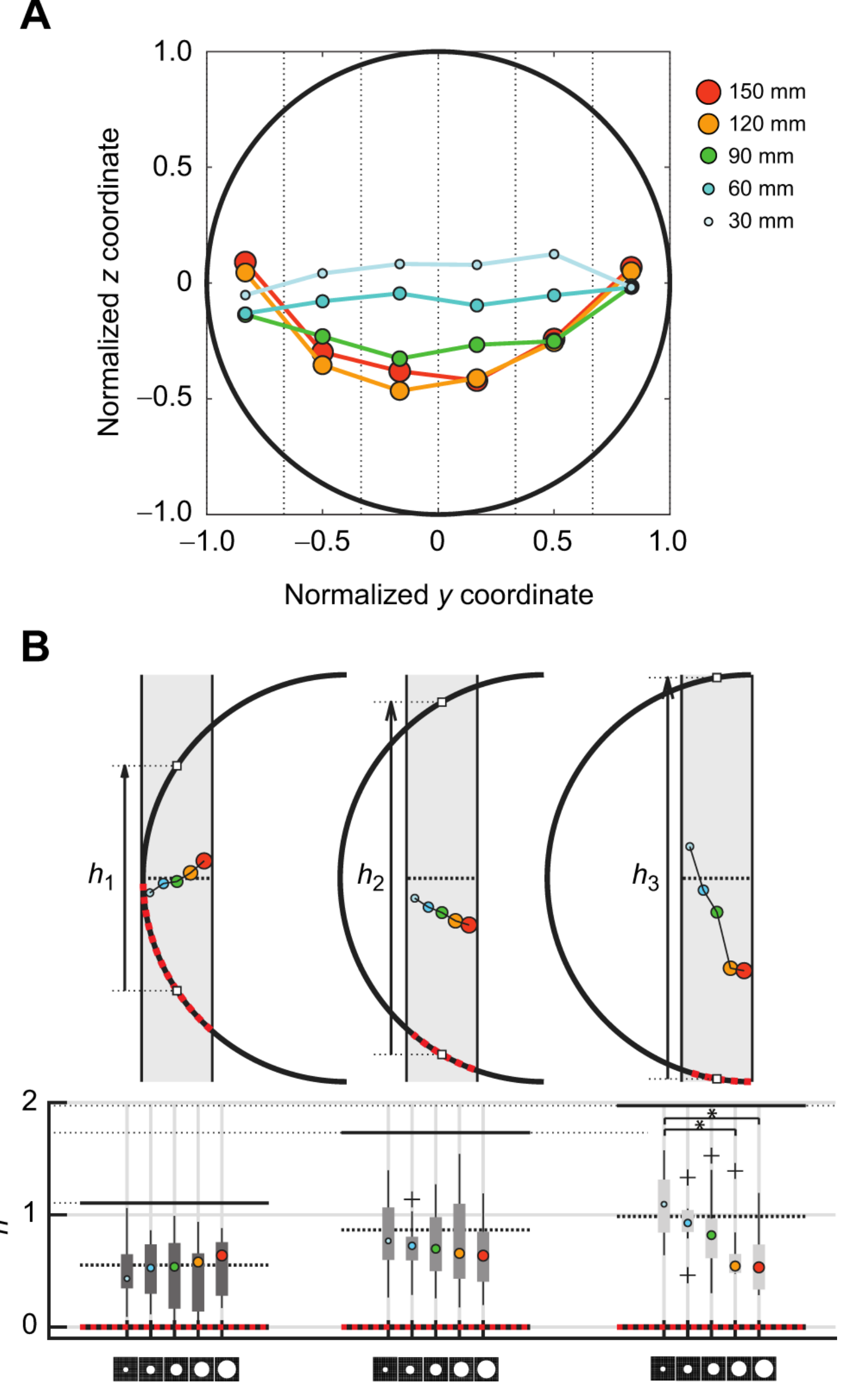

**Fig. 6. Entry heights above the lower edge for circular apertures (30–150 mm diameter), grouped into laterally spaced bins.**

(A) Diagram illustrating the bin geometry across the full aperture, normalized so that its radius is 1 along both the Y- and Z- axes. Heights are measured vertically from each entry point to the intersection with the lower aperture edge. Median height coordinates of the entry points in each of the six lateral bins are shown by circular markers and connected by colored solid lines for each aperture diameter. The size and colour of these markers correspond to the circular aperture diameters as shown in the legend on the right.

(B) Schematic showing the aperture trisectors containing the lower edge (marked by a dotted red line), the vertical height at the trisector centres ($h_{1-3}$) (shown by white square markers), and the vertical centre of the section (shown by a horizontal black dotted line). Below this, box-and-whisker plots of the normalized height above the lower edge ($h_E^*$) for all apertures (30-150 mm) are separated into the three laterally spaced bins (outer, middle and inner) corresponding to the schematic above. The colored circular markers correspond to the median height for each aperture, while three different shades of grey (darker grey for outer bins, lighter grey for inner bins) distinguish the outer and inner bins. Solid horizontal lines mark the normalized section height measured in each bin geometry ($h_{1-3}^*$) as diagrammed in the schematic above. Significance was tested across these datasets, with significant differences indicated by asterisks above the respective boxplots.

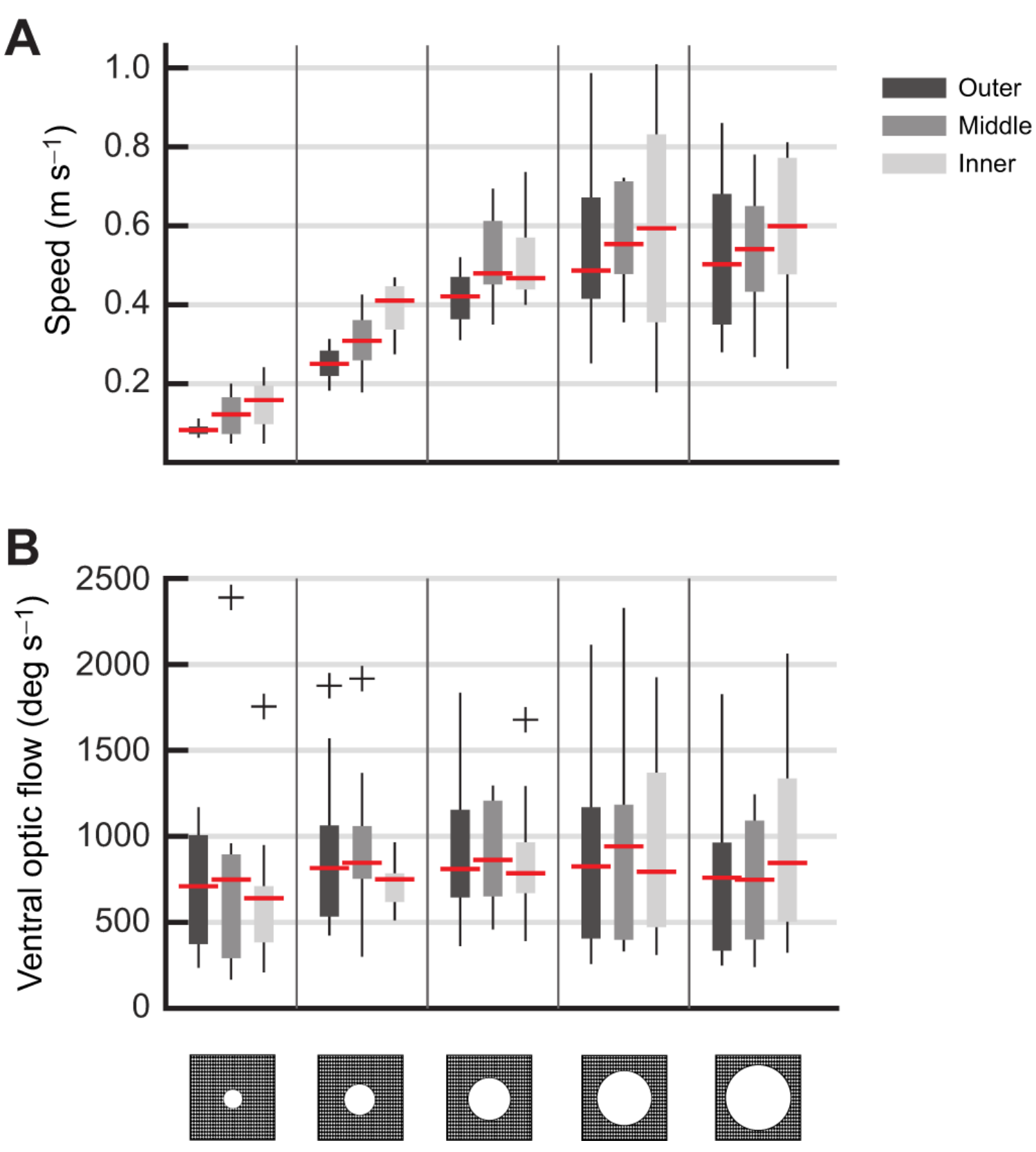

**Fig. 7. Bee traversal speed ($v_x$) and ventral optic flow ($\Omega_V$) for circular apertures in each lateral bin.**

(A) Traversal speed grouped by the bees' lateral trisector of entry ('Inner', 'Middle', and 'Outer') as diagrammed in Fig. 6B. The three different shades of grey of the box-and-whisker plots indicate the trisector they represent (darker grey for outer bins, lighter grey for inner bins) as shown in the legend on the right side. This shading corresponds to the data for each trisector shown in Fig. 6B.

(B) Ventral optic flow grouped by the bees' lateral trisector of entry. The boxplots show the median (red line), interquartile range, and any outliers, with corresponding aperture type icons below the plots (circular apertures ranging 30-150 mm diameter).

## Tables

**Table 1:** Spatial statistics for the position samples from of all tested shapes and sizes shown in Fig. 3. Aperture widths (in mm), the number of samples recorded, the median horizontal ($\tilde{Y}$) and vertical ($\tilde{Z}$) coordinates (in mm) are displayed for each aperture type (A–K). The table provides the MAD for Y and Z coordinates, the number of collisions recorded, the collision ratio (the inverted success rate, $N_{Collisions} / N_{Entries}$), and the wing strike ratio ($N_{Collisions} / N_{Wing\ Collisions}$) for each aperture.

| Label | A | B | C | D | E | F | G | H | I | J | K |
|---|---|---|---|---|---|---|---|---|---|---|---|
| Type | | | | | | | | | | | |
| Width (mm) | 30 | 60 | 90 | 120 | 150 | 120 | 60 | 60 | 60 | 60 | 60 |
| Samples | 109 | 534 | 385 | 456 | 455 | 260 | 236 | 186 | 211 | 207 | 230 |
| Median horizontal coordinate $\tilde{Y}$ (mm) | 0.21 | 0.17 | 2.51 | 1.25 | 0.95 | 2.21 | -1.20 | -0.12 | 0.44 | -2.07 | 2.89 |
| Median vertical coordinate $\tilde{Z}$ (mm) | 0.24 | -1.87 | -11.42 | -23.85 | -24.09 | -9.88 | -12.31 | 0.91 | -3.14 | -0.91 | -0.32 |
| $MAD_Y$ (mm) | 5.18 | 6.26 | 7.63 | 11.61 | 13.09 | 15.27 | 12.13 | 11.89 | 11.15 | 9.89 | 10.81 |
| $MAD_Z$ (mm) | 3.68 | 5.47 | 12.52 | 15.78 | 20.93 | 9.25 | 19.39 | 7.84 | 7.00 | 9.17 | 8.66 |
| Collisions | 67 | 149 | 70 | 55 | 46 | 25 | 120 | 98 | 94 | 97 | 96 |
| Collision ratio (inverse of success rate) | 0.615 | 0.279 | 0.182 | 0.121 | 0.101 | 0.096 | 0.508 | 0.527 | 0.446 | 0.469 | 0.417 |
| Wing strike ratio | 0.967 | 0.981 | 0.513 | 0.429 | 0.177 | 0.332 | 0.938 | 0.770 | 0.898 | 0.885 | 0.888 |

**Table 2.** Median comparisons for speed and OF metrics from all tested aperture types. The table presents the median, coefficient of variation based on the median ($CV_{Med}$), median absolute deviation (MAD), interquartile range (IQR), range, and group comparison (KW) p-values for each optic flow metric: speed ($v_x$), bilateral OF ($\Omega_{BL}$), dorsoventral OF ($\Omega_{DV}$), total radial OF ($\Omega_R$), and ventral OF ($\Omega_V$). Statistical significance of differences across aperture types is indicated by the group-level (KW) p-values in the final column. Group comparisons in the stadium category include comparisons to circular apertures of the same major and minor-axis lengths (60 mm and 120 mm circular apertures). The detailed statistics of the datasets making up these groups can be found in Table S2. Details of which groups are via post-hoc tests are listed in Table S3 and illustrated in Fig. 2.

| Comparison group | Metric | n | Median | IQR | MAD | $CV_{Med}$ | Group-Level Comparison (p) |
|---|---|---|---|---|---|---|---|
| A-E | $\Omega_{BL}$ | 116 | 765.1 | 333.6 | 223.5 | 0.292 | 0.0002 |
| | $\Omega_{DV}$ | 118 | 744.7 | 278.2 | 176.7 | 0.237 | 0.0001 |
| | $\Omega_R$ | 116 | 614.7 | 256.7 | 153.4 | 0.250 | 0.0002 |
| | $\Omega_V$ | 111 | 767.4 | 269.0 | 174.0 | 0.227 | 0.1241 |
| F, D, G, B | $\Omega_{BL}$ | 95 | 686.5 | 522.1 | 284.7 | 0.415 | <0.0001 |
| | $\Omega_{DV}$ | 94 | 717.7 | 353.0 | 227.6 | 0.317 | <0.0001 |
| | $\Omega_R$ | 93 | 619.3 | 335.4 | 210.6 | 0.340 | 0.0023 |
| | $\Omega_V$ | 79 | 784.4 | 368.8 | 218.8 | 0.279 | <0.0001 |
| B, H-K | $\Omega_{BL}$ | 112 | 842.1 | 298.4 | 210.3 | 0.250 | 0.0642 |
| | $\Omega_{DV}$ | 109 | 802.1 | 306.2 | 184.0 | 0.229 | 0.5839 |
| | $\Omega_R$ | 124 | 677.4 | 396.1 | 219.8 | 0.324 | 0.7628 |
| | $\Omega_V$ | 114 | 681.7 | 322.6 | 203.4 | 0.298 | 0.2432 |

**Table 3:** Comparison of median differences in Y and Z coordinates between the control (B) and texture-flanked apertures (H–K), as well as between apertures flanked with texture on opposite sides in the vertical and horizontal directions. The table presents the group comparison (KW) H-statistic and corresponding p-values, along with significant pairwise comparisons from Dunn's post-hoc tests ($p < 0.05$). There are two degrees of freedom for the group comparisons.

| Comparison | Metric | n | H-statistic | Group-Level Comparison (p) | Effect Size ($\epsilon^2$) | Significant Pairwise Comparisons (p) |
|---|---|---|---|---|---|---|
| B, H-I | Y coordinate | 637 | 0.57 | 0.7502 | 0.00 | All NS |
| | Z coordinate | | 4.55 | 0.1030 | 0.00 | All NS |
| B, J-K | Y coordinate | 635 | 10.49 | 0.0053 | 0.02 | J vs K (p = 0.0036) |
| | Z coordinate | | 4.67 | 0.0969 | 0.01 | All NS |

## Supplementary Information

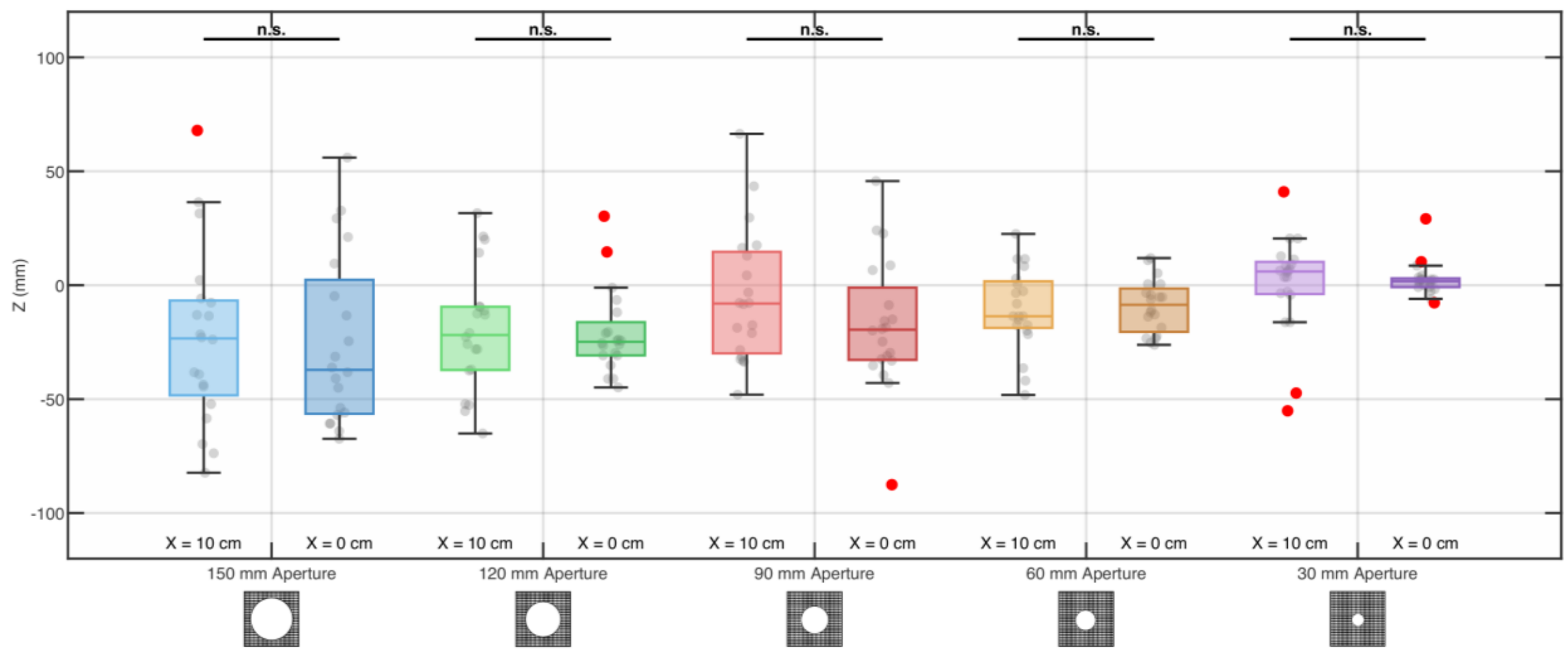

**Figure S1:** Comparison of the bee Z coordinates for each of the circular apertures (30-150 mm) at two sampling planes: one at 10 cm before reaching the transverse wall (X = 10 cm) and one at the location of traversal (X = 0 cm). Datasets were compiled by analysing 20 randomly chosen bee trajectories in each case. Samples from each dataset are plotted behind each corresponding box plot, with outliers displayed as red markers. Paired Wilcoxon signed-rank tests detected no significant change in height between the two sampling planes for all aperture sizes (n.s.).

**Table S1:** Paired Wilcoxon signed-rank tests comparing height (Z) at X = 0 cm (traversal plane) and X = 10 cm (10 cm before the aperture wall traversal plane) for each circular aperture diameter. $\Delta Z$ = median paired difference ($Z_0 - Z_{10}$) in mm; IQR = interquartile range of $\Delta Z$; W = sum of positive ranks with zero differences omitted; Wilcoxon effect size, r = $\frac{|Z|}{sqrt(n)}$. All comparisons are non-significant (p>0.05) and conclusions were unchanged after holm-correction across the five tests.

| Aperture | n | $\Delta Z$ ($Z_0 - Z_{10}$), mm | IQR, mm | Test statistic, W | p (two-tailed) | Effect size, r |
|---|---|---|---|---|---|---|
| | 20 | -9.20 | 30.3 | 122 | 0.5257 | 0.14 |
| | 20 | -2.74 | 21.8 | 119 | 0.6012 | 0.12 |
| | 20 | -11.70 | 53.1 | 149 | 0.1005 | 0.37 |
| | 20 | 1.75 | 17.1 | 94 | 0.6813 | 0.09 |
| | 20 | -2.45 | 22.8 | 108 | 0.9108 | 0.03 |

**Table S2:** Summary of visual flow statistics for all apertures of tested. The table includes the median speed ($V_X$), interquartile range (IQR), median absolute deviation (MAD), and coefficient of variation based on the median ($CV_{Med}$) for each aperture size (A–E). Group-level comparison p-values (KW test outputs) are included to indicate the statistical significance of differences across aperture sizes.

| Label | Type | Metric | n | Median | IQR | MAD | $CV_{Med}$ | Group-Level Comparison (p) |
|---|---|---|---|---|---|---|---|---|
| A | | | 24 | 718.8 | 244.1 | 174.3 | 0.242 | |
| B | | | 24 | 897.6 | 379.0 | 225.8 | 0.252 | |
| C | | $\Omega_{BL}$ | 21 | 799.8 | 494.4 | 248.3 | 0.310 | 0.0002 |
| D | | | 23 | 794.6 | 365.5 | 226.6 | 0.285 | |
| E | | | 24 | 636.3 | 161.6 | 155.7 | 0.245 | |
| A | | $\Omega_{DV}$ | 24 | 648.4 | 345.7 | 188.6 | 0.291 | 0.0001 |
| B | | | 25 | 840.7 | 173.8 | 113.3 | 0.135 | |
| C | | | 23 | 847.7 | 261.1 | 171.2 | 0.202 | |
| D | | | 22 | 744.7 | 292.4 | 182.7 | 0.245 | |
| E | | | 24 | 622.8 | 255.1 | 145.2 | 0.233 | |
| A | | $\Omega_R$ | 24 | 492.8 | 286.0 | 166.7 | 0.338 | 0.0002 |
| B | | | 22 | 694.1 | 205.9 | 125.1 | 0.180 | |
| C | | | 23 | 675.1 | 214.6 | 114.4 | 0.169 | |
| D | | | 24 | 590.1 | 229.9 | 152.1 | 0.258 | |
| E | | | 23 | 533.2 | 262.6 | 135.8 | 0.255 | |
| A | | $\Omega_V$ | 21 | 665.6 | 226.3 | 136.7 | 0.205 | 0.1241 |
| B | | | 24 | 763.5 | 243.8 | 151.1 | 0.198 | |
| C | | | 24 | 820.5 | 417.7 | 189.1 | 0.230 | |
| D | | | 22 | 827.5 | 318.8 | 187.5 | 0.227 | |
| E | | | 20 | 765.3 | 335.9 | 181.5 | 0.237 | |
| F | | $\Omega_{BL}$ | 23 | 320.1 | 158.4 | 118.3 | 0.370 | <0.0001 |
| G | | | 25 | 595.3 | 530.7 | 275.6 | 0.463 | |
| F | | $\Omega_{DV}$ | 22 | 767.5 | 234.7 | 200.4 | 0.261 | <0.0001 |
| G | | | 25 | 336.0 | 240.7 | 131.6 | 0.392 | |
| F | | $\Omega_R$ | 22 | 764.7 | 519.7 | 339.5 | 0.444 | 0.0023 |
| G | | | 25 | 450.1 | 163.8 | 136.9 | 0.304 | |
| F | | $\Omega_V$ | 16 | 916.0 | 94.2 | 134.9 | 0.147 | <0.0001 |
| G | | | 17 | 379.1 | 432.3 | 198.5 | 0.523 | |
| H | | $\Omega_{\mathrm{BL}}$ | 22 | 707.3 | 391.0 | 231.2 | 0.327 | 0.0642 |
| I | | | 22 | 930.7 | 330.6 | 228.4 | 0.245 | |
| J | | | 23 | 824.2 | 399.8 | 201.5 | 0.244 | |
| K | | | 21 | 805.3 | 218.8 | 126.4 | 0.157 | |
| H | | $\Omega_{DV}$ | 20 | 706.2 | 362.8 | 203.8 | 0.289 | 0.5839 |
| I | | | 21 | 790.6 | 391.0 | 224.3 | 0.284 | |
| J | | | 21 | 816.9 | 407.1 | 214.2 | 0.262 | |
| K | | | 22 | 752.9 | 302.0 | 174.4 | 0.232 | |
| H | | $\Omega_R$ | 23 | 657.7 | 349.7 | 208.5 | 0.317 | 0.7628 |
| I | | | 26 | 707.2 | 430.9 | 274.8 | 0.389 | |
| J | | | 26 | 708.0 | 489.6 | 258.7 | 0.365 | |
| K | | | 27 | 645.7 | 421.0 | 216.7 | 0.336 | |
| H | | $\Omega_V$ | 22 | 634.3 | 463.7 | 233.8 | 0.369 | 0.2432 |
| I | | | 23 | 696.8 | 251.0 | 196.9 | 0.283 | |
| J | | | 22 | 636.1 | 170.2 | 155.0 | 0.244 | |
| K | | | 23 | 636.9 | 259.0 | 494.4 | 0.407 | |

**Table S3:** Group comparison test results for circular aperture group comparisons across various OF metrics. The table displays the number of samples (n), H-statistic ($\chi^2$), KW p-values for the overall comparison of groups, and significant Dunn's post-hoc pairwise comparisons for speed, bilateral OF ($\Omega_{BL}$), dorsoventral OF ($\Omega_{DV}$), radial OF ($\Omega_R$), and ventral OF ($\Omega_V$). Significant Dunn comparisons are indicated with asterisks: $p < 0.05$ (*), $p < 0.01$ (**), and $p < 0.001$ (***). NS indicates non-significant results. There are four degrees of freedom.

| Comparison | Metric | n | H-Statistic ($\chi^2$) | Group-Level Comparison (p) | Effect size ($\eta$) | Significant Pairwise Comparisons (p) |
|---|---|---|---|---|---|---|
| A-E | $\Omega_{BL}$ | 116 | 22.3 | 0.0002 | 0.16 | B vs E (p = 0.0015)**<br>C vs E (p = 0.0017)** |
| | $\Omega_{DV}$ | 118 | 22.9 | 0.0001 | 0.17 | A vs B (p = 0.0474)*<br>A vs C (p =  0.0101)*<br>B vs E (p = 0. 0067)**<br>C vs E (p = 0.0011)** |
| | $\Omega_R$ | 116 | 22.3 | 0.0002 | 0.17 | A vs B (p = 0.0335)*<br>A vs C (p = 0.0045)**<br>B vs E (p = 0.0214)*<br>C vs E (p = 0.0027)** |
| | $\Omega_V$ | 111 | 7.23 | 0.1241 | 0.03 | NS |
| F, D, G, B | $\Omega_{BL}$ | 95 | 44.2 | <0.0001 | 0.45 | F vs D (p = 0.0000)***<br>F vs G (p = 0.0010)**<br>F vs B (p = 0.0000)***<br>G vs B (p = 0.0500)* |
| | $\Omega_{DV}$ | 94 | 50.5 | <0.0001 | 0.53 | F vs G (p = 0.0000)***<br>D vs G (p = 0.0000)***<br>G vs B (p = 0.0000)*** |
| | $\Omega_R$ | 93 | 14.49 | 0.0023 | 0.13 | F vs G (p = 0.0039)**<br>G vs B (p = 0.0241)* |
| | $\Omega_V$ | 79 | 29.8 | <0.0001 | 0.36 | F vs G (p = 0.0000)***<br>D vs G (p = 0.0006)***<br>G vs B (p = 0.0020)** |
| B, H-K | $\Omega_{BL}$ | 112 | 8.88 | 0.0642 | 0.05 | NS |
| | $\Omega_{DV}$ | 109 | 2.85 | 0.5839 | -0.01 | NS |
| | $\Omega_R$ | 124 | 1.85 | 0.7628 | -0.02 | NS |
| | $\Omega_V$ | 114 | 5.46 | 0.2432 | 0.01 | NS |

Movie Legends:

**Supplementary Movie 1: FPV flythrough along a representative bee trajectory through the flight tunnel.** The video displays the scene from the bee's perspective (rendered at 90 deg FOV for human viewing), with key tunnel features along the flight path annotated. The virtual camera was rotated in tilt, pan and yaw across the trajectory so that its optical axis remained aligned with the X axis of the tunnel, mimicking the gimballed head stabilisation of a flying bee between saccadic gaze shifts.

**Supplementary Movie 2: Retinal velocity heatmap of a simulated wide-angle first-person-view flight trajectory through a 120 mm aperture.** The simulation provides a visual approximation of the translational optic flow as the bee traverses the aperture. The virtual camera was rotated in tilt, pan and yaw across the trajectory so that its optical axis remained aligned with the X axis of the tunnel, mimicking the gimballed head stabilisation of a flying bee between saccadic gaze shifts. Raw frames from the simulation were processed with a dense Farnebäck optic flow algorithm to derive per-pixel motion vectors. Magnitudes were smoothed with a five-frame rolling average and clipped at the $90^{th}$ percentile to supress outliers. The resulting magnitudes were normalised to the global maximum computed across all frames and color-coded accordingly (cooler hues = low speeds; warmer hues = high speeds). Playback is at the native frame rate of 60 hz, with each frame rendered at 960 x 540 px. The uniform checkerboard pattern introduces spatial aliasing in the rendered heat map when it is distant from the camera. This artefact is expected: as distance decreases, the projected spatial frequency eventually exceeds the Nyquist limit. Insects are largely immune to this effect due to the inter-ommatidial spacing of a bee’s compound eye (1–2 deg) low-pass-filtering fine patterns, and irregular receptor mosaics further suppressing coherent aliasing.

**Supplementary Movie 3: Example bee traversal without a collision.**
A typical video recording of a bee traversal in which no collision occurred. The view field shows the front view of the transverse wall (YZ plane), looking longitudinal down the tunnel from just below the entrance. A honeybee can be seen approaching the aperture plane. At the moment of traversal, the video freezes and highlights the bee’s location using a green locating circle and a red dot. The mirror view (facing the XY plane) shows when the bee intersects the aperture plane, and a corresponding green locating circle and a red dot is placed to show this location.

**Supplementary Movie 4: Example bee traversal with a collision.**

A typical video recording of a bee traversal when a collision occurred. A honeybee can be seen approaching the aperture plane and experiencing a body/leg collision with the lower edge of the aperture. At the moment of traversal, the video freezes and highlights the bee's collision location on the rim using a green locating circle and a red square in both views. This process of marking coordinates as shown was used to collect the entry and collision coordinates of all flights.